\documentclass{amsart}
\usepackage{amssymb}
\usepackage{amsfonts}
\usepackage{amsmath}
\usepackage[legalpaper,bookmarks=true,colorlinks=true,linkcolor=blue,citecolor=blue]{hyperref}
\usepackage{graphicx}%
\usepackage{fancyhdr}
\usepackage{color}
\usepackage[mathlines]{lineno}
\usepackage{lscape}
\usepackage{epsfig}
\usepackage{natbib}
\usepackage{geometry}
\usepackage{tgbonum}
\usepackage[]{listings}

\fontfamily{qcr}\selectfont

\newtheorem{theorem}{Theorem}
\theoremstyle{plain}

\newtheorem{proposition}{Proposition}

\numberwithin{equation}{section}

\newcommand{\Bin}{\bigskip \noindent}

\newcommand{\Ni}{\noindent}

\begin{document}
\Large
\title[Two samples problems under iid date]{Asymptotic Statistical Theory for the Samples Problems using the Functional Empirical Process, revisited I}
\normalsize
\author{Abdoulaye Camara $^{(1)}$, Adja Mbarka Fall $^{(2)}$, Moumouni Diallo $^{(3)}$, Gane Samb Lo $^{(4)}$}

\maketitle
\Ni\textbf{Abstract}. In this paper we study the asymptotic theory for samples problem based on the functional empirical process (fep), this new method is called general samples problem. We suggest this method to develop the full theory of estimation of means, variances, ratios of variances and difference of means for independent samples. We compare the results of our new method to the Gaussian method using simulated and real data. The obtained results are almost equivalent to those in the Gaussian case for samples's size equal to $10$. It has been prove that the estimation of the means difference is very precise regardless of the equality or inequality of variances for greater sizes of sample. This method is recommended when the sizes of samples is around or greater that $15$ and it requires the finiteness of the fourth order moment.\\

\Ni \textbf{NB}. Gane Samb Lo is not author in the first version which published in a peer-reviewed journal, although his took a main part in the investigation. The current version archived in arxiv.org hols all authors and this version will revisited to inlude more results and data-driven applications.

\newpage
\Ni $^{(1)}$ Abdoulaye Camara\\
Universit\'e des Sciences, Techniques et des Technologies, Bamako.\\
LERSTAD, Gaston Berger University (UGB), Saint-Louis, Senegal.\\
IMHOTEP International Mathematical Centre (IMHO-IMC: imhotepsciences.org)\\
\noindent $^{(2)}$ Dr Moumouni Diallo\\
Université des Sciences Sociale et de Gestion de Bamako ( USSGB)\\
Faculté des Sciences Économiques et de Gestion (FSEG)\\
IMHOTEP International Mathematical Centre (IMHO-IMC: imhotepsciences.org)\\
Email: moudiallo1@gmail.com.\\
\noindent $^{(3)}$ Dr. Adja Mbarka Fall\\
Unversité Iba Der Thiam de  Thi\`es, SENEGAL\\
IMHOTEP International Mathematical Centre (IMHO-IMC: imhotepsciences.org)\\
\noindent $^{\dag}$ Gane Samb Lo.\\
LERSTAD, Gaston Berger University, Saint-Louis, S\'en\'egal (main affiliation).\newline
LSTA, Pierre and Marie Curie University, Paris VI, France.\newline
AUST - African University of Sciences and Technology, Abuja, Nigeria\\
IMHOTEP International Mathematical Centre (IMHO-IMC: imhotepsciences.org)\\
gane-samb.lo@edu.ugb.sn, gslo@aust.edu.ng, ganesamblo@ganesamblo.net\\
Permanent address : 1178 Evanston Dr NW T3P 0J9, Calgary, Alberta, Canada.\\

\newpage
\Ni\textbf{Presentation of authors}.\\

\noindent \textbf{Abdoulaye Camara}, M.Sc., is an assistant professor of Mathematics and Statistics at: Universit\'e des Sciences, Techniques et des Technologies, Bamako. He is preparing a PhD thesis at: LERSTAD, Unversit\'e Gaston Berger Saint-louis, Senegal. He is also a junior researcher at: Mathematical Center (imho-imc), https://imhotepsciences.org.\\

\noindent \textbf{Adja Mbarka Fall}, PhD, is an Associated Professor of Mathematics and Statistics at: Universit\'e Iba Der Thiam de Thies, SENEGAL. She is also a senior researcher at: Mathematical Center (imho-imc), https://imhotepsciences.org.\\

\noindent \textbf{Moumouni Diallo}, PhD, is a Professor of Mathematics and Statistics at:Universit\'e des Sciences Sociales et de Gestion, Bamako, MALI. He is a senior researcher at: Mathematical Center (imho-imc), https://imhotepsciences.org.\\

\Ni \textbf{Gane Samb Lo}, Ph.D., is a retired full professor from Université Gaston Berger (2023), Saint-Louis, SENEGAL. He is the founder and The Probability and Statistics Chair holder at: Imhotep International Mathematical Center (imho-imc), https://imhotepsciences.org.\\
\newpage

\section{Introduction} \label{sec_01}

\Ni In general, the statistical theory for linear methods in particular analysis of variance (ANOVA) and linear regression is mainly based on the Gaussian assumption. That theory of Gaussian linear models is generally well documented with a large number of teaching books and research books. At least we may cite a few including \cite{seber}, \cite{ScheffeAV}, \cite{tibshirani}, etc.\\
\Ni In that frame, the theory works fine. However in many situations the data are not Gaussian. For example accidental data are more likely to follow the gamma law. It happens that the drift from the Gaussian hypothesis can not lead to a so beautiful theory. Let us consider two types of sets.\\

\Ni (a) In \cite{ScheffeAV}, we pick the data as in page \pageref{scheffe}. Therein, we have measurements of waterproof quality of sheets of material manufactured on five machines on nine days. Here, we have three batches of data. In each batch, lines refer to data from machine A, B and C. The study here is whether or not a machine gives more robust sheets that another in the sens of higher quality of waterproof.  The normality Jarque-Bera test works fine for all variables (taken horizontally) and their differences, as we checked it in Subsection \ref{sec_05_01} of Section \ref{sec_05}. For these data, the Gaussian methods are relevant and may be used.\\ 

\Ni (b) Income data from database ESAM1 (1996) and ESAM2 (2000) are considered. For regions Dakar and Diourbel, the income data are dakar1 and dakar2, diour1 and diour2. In this study, we consider a sampling of $T$ observation from each set of data, just for illustration. We want to compare the the revenue between the two areas for each survey and eventually to try to see the evolution for one region, from one survey to the other. As we will see it just in the subsection mentioned above,  these data are clearly non Gaussian.\\

\Ni In situations like (b), it is not reasonable to use the Gaussian method. As we already mentioned, there is a significant number of areas in which Gaussian data are not expected. A clear example concerns heavy tails data or mixing data in which they are involved.\\

\Ni In such situations, we propose to use asymptotic methods in the spirit of \cite{loFepLrepMethod} through asymptotic representations of the involved statistics. That method is a consequence of the whole theory of weak convergence. In the above cite article,  a full list of related works and books on weak convergence can be found.\\

\Ni Our best achievement is the setting of full theory of estimation of means, variances, ratios of variances and differences of means for independent samples with sizes around or greater that $n\geq 15$. The asymptotic laws needed for building the confidence intervals are given in Theorems \ref{gmOS} (page \pageref{gmOS}) and \ref{gmTIS} (page \pageref{gmTIS}). Those confidence intervals are displayed Subsection \ref{sec_05_01} (page \pageref{sec_05_01}) for the one sample problem and Subsection \ref{sec_05_06} (page \pageref{sec_05_06}) for the two independent sample problem.\\

\Ni The rest of the paper is organized as follows. We begin by recalling the nice Gaussian theory in Section \ref{sec_02} while we give our results in Section \ref{sec_03} where a full theory of comparison based on the empirical process for general data. We focus on the implementation of the results Section \ref{sec_04}. Next the results are simulated and applied to datadriven examples in Section \ref{sec_05} where a number of scenario are presented and discussed. We close the paper with Section \ref{sec_06} as the paper conclusion.\\

\section{The Gaussian Samples problems} \label{sec_02}

\subsection{Introduction}

\Ni The samples problems for Gaussian data are treated in a parametric frame and use exact distribution theory. It is quite impossible to have a complete theory for other distributions. In order to able to highlight the results obtained from the use of the \textit{fep}, we are going to recall the striking distribution facts for Gaussian samples.\\

\bigskip \noindent This part introduces to the standard statistical methods related to Gaussian samples. It comes after the study of Gaussian vectors, of which it presents statistical applications. It is a preparation to the wider field of linear regression, analysis of variance (ANOVA) and analysis of covariance (ANCOVA). We will focus on the one sample and the two sample problems.\\

\noindent The techniques given here, in despite of their simplicity, yield very powerful methods in a number of fields, including Life Sciences, Medical and Pharmaceutical Sciences, Industrial Studies, etc., in short, in any discipline in which data may be modeled as Gaussian.\\

\subsection{The Gaussian sample problem techniques} \label{sec1}

\noindent A Gaussian sample is simply of size $n$ a sequence $X_{1},X_{2},...,X_{n}$ of $n$ independent and identically distributed (\textit{iid}) random variables with a Gaussian common probability law of mean $m$ and $\sigma^2$ denoted

$$
(H) \qquad X_{1},X_{2},...,X_{n} \ \sim \mathcal{N}(m,\sigma^2).
$$

\bigskip \noindent We always suppose that $n\geq 2$, otherwise doing statistics is meaningless. The two main statistics are the sample mean:

\begin{equation}
\overline{X}_{n}=\frac{1}{n}\sum_{i=1}^{n}X_{i}  \label{sm.eg.mean},
\end{equation}

\noindent and the  sample variance

\begin{equation}
S_{n}^{2}=\frac{1}{n-1}\sum_{i=1}^{n}\left( X_{n}-\overline{X}_{n}\right)^{2}.  \label{sm.eg.varemp}
\end{equation}

\subsection{The one Gaussian sample problem}

\noindent Here is the most important result concerning the distributional properties of the Gaussian sample

\begin{proposition} \label{fact_01} For all $n\geq 2$, we have :\\

\noindent (a) $S_{n}^{2} \qquad  \text{is independent of} \qquad \overline{X}_{n}$\\

\noindent (b) \textit{Law of \quad   $\overline{X}_{n}$, when $\sigma$ is known}.

\begin{equation}
\frac{\sqrt{n}\left( \overline{X}_{n}-m\right) }{\sigma}\sim \mathcal{N}\left( 0,1\right)  \label{sm.eg.06}
\end{equation}

\noindent (c) \textit{Law of \quad  $S_n^2$}.

\begin{equation}
\left( n-1\right) S_{n}^{2}/\sigma ^{2}\sim \chi _{n-1}^{2}  \label{sm.eg.05}
\end{equation}

\noindent (d) \textit{Law of \quad $\overline{X}_{n}$, when $\sigma$ is unknown}.

\begin{equation}
\frac{\sqrt{n}\left( \overline{X}_{n}-m\right) }{S_n} \sim t(n-1).   \label{sm.eg.04}
\end{equation}
\end{proposition}

\subsection{Two independent Gaussian samples problems}

\Ni We have a first easy case where the two data to be compared are linked to the same statistical unit. \\

\Ni {A - Two paired sample Gaussian problem}.\\

\noindent Suppose that we have paired measures $\left( X_{i},Y_{i}\right) $ associated with the same individuals, $i=1,...,n$. For example, for the same person, $X$ might represent blood sugar and $Y$ the cholesterol measure.\\

\noindent Assume that the $\left( X_{1},Y_{1}\right) ,\left( X_{2},Y_{2}\right) ,...,\left( X_{n},Y_{n}\right) $ are independent Gaussian random couples such that

$$
(H0) \qquad  \forall \quad  i\leq i\leq n, \ \mathbb{E}X_{i}=m_{1} \textit{ and } \mathbb{E}Y_{i}=m_{2}
$$

\noindent We want to test the hypothesis

$$
(H) \quad   m_{1}=m_{2},
$$

\noindent which is  equivalent to test,  $\Delta{m}=m_{1}-m_{2}=0$.\\

\noindent Put 
$$
Z_{i}=X_{i}-Y_{i},  \quad i=1,\dots,n.
$$

\noindent Given that the pair $(X_1,Y_2)$ is Gaussian, we can and do transform the two samples problem a one sample problem of the data $Z_{1},...,Z_{n}$ of common probability law $\mathcal{N}\left( \Delta _{m},\sigma^{2}\right)$, where $\sigma^{2}=\mathbb{V}(X_1-X_2)$.\\

\Bin Now, the most interesting case is the following.\\
 
\Ni \textbf{B - The two Gaussian independent samples}.\\

\noindent Suppose we have  

\begin{equation*}
\mathcal{E}_X : \ X_{1},X_{2},...,X_{n} \quad \text{ iid } \sim \mathcal{N}\left( m_{1},\sigma_{1}^{2}\right),
\end{equation*}

\bigskip \noindent and
\begin{equation*}
\mathcal{E}_Y : \  Y_{1},Y_{2},...,Y_{m} \quad \text{ iid } \sim \mathcal{N}\left( m_{2},\sigma_{2}^{2}\right).
\end{equation*}

\Bin \textbf{B1 - The law of the ratio of variances.}\\

\Ni We denote by $S_n^2$ and $S_m^2$ the empirical variances for the samples $\mathcal{E}_X$ and $\mathcal{E}_Y$ respectively. We consider the ratio of variances $R=\sigma_1^2/ \sigma_2^2$. 

\begin{proposition} \label{fact_02} We have
\begin{equation*}
\frac{1}{R^{2}}\frac{S_{n}^{2}}{S_{m}^{2}} \sim F_{n,m},
\end{equation*}

\Bin where $F_{n,m}$ stands for the Fisher law of degrees of freedom $n$ and $m$.
\end{proposition}

\Bin \textbf{B2 - Estimation of $\Delta m$}.\\

\Ni The estimation is done in two cases.\\

\Ni \textbf{B2a - Case of equal variances}.\\

\noindent Denote $\sigma _{1}^{2}=\sigma _{2}^{2}=\sigma ^{2}$ and

\begin{equation*}
S^{2}=\frac{\left( n-1\right) S_{n}^{2}+\left( m-1\right) S_{m}^{2}}{n+m-2}.
\end{equation*}

\begin{proposition} \label{fact_03} We have

\begin{equation*}
\frac{\overline{X}_{n}-\overline{Y}_{m}-\Delta m}{s\sqrt{\frac{1}{n}+\frac{1%
}{m}}}\sim t(n+m-2).
\end{equation*}
\end{proposition}

\Bin \textbf{B2b - Case of unequal variances}\\

\bigskip \noindent In the literature, in such a case, the approximation of \textit{Aspin-Welch test} and \textsl{Welch's t-test} (Welch, 1937) using the \textit{Satterthwaite} method is used as follows:\\ \label{aspin}
 
\begin{proposition} \label{fact_4}

\begin{equation*}
\frac{\overline{X}_{n}-\overline{Y}_{m}-\Delta m}{\sqrt{\frac{S_{n}^{2}}{n}+\frac{S_{m}^{2}}{m}}}\sim t(f),
\end{equation*}

\bigskip \noindent where
\begin{equation*}
f=\frac{\left( \frac{S_{n}^{2}}{n}+\frac{S_{m}^{2}}{m}\right) ^{2}}{\frac{1}{%
n-1}\left( \frac{S_{n}^{2}}{n}\right) ^{2}+\frac{1}{m-1}\left( \frac{%
S_{m}^{2}}{m}\right) ^{2}}.
\end{equation*}
\end{proposition}

\Bin \textbf{Remark}. This fascinated and most used tool for Gaussian data is justified in our simulation works below. We still think that a proof at the level of the modern weak convergence theory is welcome. We do think that the approached developed in this paper is the right one.

\section{Alternative theory using the fep} \label{sec_03}

\Bin In this section, we are concerned with asymptotic theory of samples problems based on the functional empirical process. \\

\subsection{The one sample problem}

\Ni Let $X, X_1, X_2, \dots$ be a sequence of independent and identically distributed (i.i.d.) random variables defined on the same probability space $(\Omega, \mathcal{A},\mathbb{P})$ with the functional empirical process (fep) $\mathbb{G}_n$. \\

\Bin We assume that 

$$
\forall i, \quad 1 \leq i \leq n,  \qquad  \mathbb{E}(X_i)=m , \qquad \mathbb{V}ar(X_i)= \sigma^2, 
$$

$$
m_k = \mathbb{E}(X^{k}), \qquad  \mu_k = \mathbb{E}\left( X - \mathbb{E}(X) \right)^k, \qquad \mu_{k,n} = \frac{1}{n} \sum_{j=1}^n \left(X_j - \overline{X}_n \right)^k,  \qquad  k \geq 1.
$$

\Bin We define the statistic $T_n^2$ given by 

\begin{equation}
T_n^2 = \mu_{4,n} - S_n^4.
\end{equation}

\Bin We have two main results relative to the mean $m$ and the variance $\sigma^2$. \\

\begin{theorem} \label{gmOS} For all $n\geq 2$, we have :\\
	
	\noindent (a) $\frac{\sqrt{n} \left(\overline{X}_{n} - m \right)}{S_n} \sim \mathcal{N}\left( 0,1\right),$\\
	
	\noindent (b)  $\frac{\sqrt{n} \left(S_n^2 - \sigma^2\right)}{T_n} \sim \mathcal{N}\left( 0,1\right).$
\end{theorem}

\Bin \textbf{Proof of Theorem \ref{gmOS}}

\Bin \textit{Proof of (a)}. Let $\alpha_n = \frac{\sqrt{n} \left(\overline{X}_{n} - m \right)}{S_n}$. By using the asymptotic representation, we have\\

\begin{equation}
	\overline{X}_{n} = m + n^{-1/2}\mathbb{G}_n(h_1),
\end{equation}

\Bin that implies that 

\begin{equation}
\overline{X}_{n} - m = n^{-1/2}\mathbb{G}_n(h_1),
\end{equation}

\begin{equation}
S_n^2 = \frac{n}{n-1} \left( \frac{1}{n} \sum_{i=1}^n X_i^2 - \overline{X}_n^2 \right) =  \frac{n}{n-1} \left( \overline{X_n^2}- \overline{X}_n^2 \right).
\end{equation}

\Bin By the delta  method, we have

\begin{equation}
\overline{X_n}^2 = m^2 + n^{-1/2} \mathbb{G}_n(2mh_1) + o_{\mathbb{P}}\left(n^{-1/2} \right),
\end{equation}

\Bin and 

\begin{equation}
	\overline{X_n^2} = m_2 + n^{-1/2} \mathbb{G}_n(h_2),
\end{equation}

\Bin where $h_j(x) = x^j$.   We know that 

\begin{equation*}
\frac{n}{n-1} = \left(\frac{n-1}{n}\right)^{-1} = \left(1 - 1/n \right)^{-1} = 1 + O_{\mathbb{P}}(1).
\end{equation*}

\Bin Hence, we have 

\begin{equation}	
S_n^2 = \sigma^2 + n^{-1/2}\mathbb{G}_n(H) + o_{\mathbb{P}} \left(n^{-1/2} \right),
\end{equation}

\Bin where $H=h_2 - 2mh_1$.  By the delta method for $S_n^2$, we get

\begin{equation}	
S_n = \sigma + n^{-1/2}\mathbb{G}_n\left(\frac{1}{\sigma}H\right) + o_{\mathbb{P}} \left(n^{-1/2} \right).
\end{equation}

\Bin By combining the asymptotic representation for $S_n$ and $\overline{X}_n - m$, we obtain

\begin{equation*}	
\frac{\overline{X}_{n} - m}{S_n} = n^{-1/2}\mathbb{G}_n\left(\frac{1}{\sigma}h_1 \right) + o_{\mathbb{P}} \left( n^{-1/2}\right) = n^{-1/2}\mathbb{G}_n\left(h \right) + o_{\mathbb{P}} \left( n^{-1/2}\right),
\end{equation*}

\Bin where $h = \frac{1}{\sigma}h_1$.  Thus 

\begin{equation}
\alpha_n =\frac{\sqrt{n} \left(\overline{X}_{n} - m \right)}{S_n} = \mathbb{G}_n\left(h\right) + o_{\mathbb{P}}(1).
\end{equation}

\Bin Hence $\alpha_n \sim \mathcal{N}\left( 0,\Gamma(h,h)\right)$,

$$
\Gamma(h,h) = \mathbb{V}ar(h(X)) =  \mathbb{V}ar\left(\frac{1}{\sigma}h_1(X)\right) = \frac{1}{\sigma^2}\mathbb{V}ar(h_1(X)) = 1.
$$

\Bin So $\alpha_n \sim \mathcal{N}\left( 0,1\right)$.\\

\Bin \textbf{Proof of (b)}.\\

\Ni Let $J_n = \frac{\sqrt{n}(S_n^2 - \sigma^2)}{\rho_n}$. We know that 

\begin{equation}
S_n^2 =\sigma^2 + n^{-1/2}\mathbb{G}_n(H) + o_{\mathbb{P}} \left(n^{-1/2} \right),
\end{equation}

\Bin where $H = h_2 - 2mh_1$. We have 

\begin{equation}
S_n^2 - \sigma^2 = n^{-1/2}\mathbb{G}_n(H) + o_{\mathbb{P}} \left(n^{-1/2} \right).
\end{equation}

\Bin The variance of $H$ is given by 

$$
\mathbb{V}ar \left(H(X)\right) = \mu_4 - \sigma^4.
$$

\Bin We also know that 

\begin{equation}
	\mu_{4,n} = \mu_4 + n^{-1/2}\mathbb{G}_n(C_0) + o_{\mathbb{P}} \left(n^{-1/2} \right),
\end{equation}

\Bin where $ C_0 = h_4 + 4m h_3 + 6m^2 h_2 + (-4m_3 + 12 m m_2 -12 m^3)h_1 $. By the delta method, we have 

\begin{equation}
S_n^4 = \sigma^4 + n^{-1/2} \mathbb{G}_n \left(2 \sigma^2 H \right) + o_{\mathbb{P}}\left( n^{-1/2}\right)
\end{equation}

\Bin From the asymptotic representation of $\mu_{4,n}$ and $S_n^4$, we get 

\begin{equation*}
T_n^2 = T^2 + n^{-1/2} \mathbb{G}_n \left(C_0 - 2\sigma^2 H \right) + o_{\mathbb{P}}\left( n^{-1/2}\right),
\end{equation*}

\Bin where  $T^2 = \mu_4 - \sigma^4$.  By the delta method, we have 

\begin{equation}
T_n = T + n^{-1/2}\mathbb{G}_n\left(L\right) + o_{\mathbb{P}}\left( n^{-1/2}\right),
\end{equation}

\Bin where  $L = C_0 - 2\sigma^2H$. From the asymptotic representation of $S_n^2 - \sigma^2$ and $T_n$, we get

\begin{equation}
\frac{S_n^2 - \sigma^2}{T_n} = n^{-1/2}\mathbb{G}_n \left( \frac{1}{T} H\right) +  o_{\mathbb{P}}\left( n^{-1/2}\right).
\end{equation}

\Bin Therefore, we have 

\begin{equation}
	\frac{\sqrt{n}\left(S_n^2 - \sigma^2\right)}{T_n} = \mathbb{G}_n \left( D\right) +  o_{\mathbb{P}}\left( 1\right),
\end{equation}

\Bin where $D = \frac{1}{T} H$. Thus 

\begin{equation*}
	\frac{\sqrt{n}\left(S_n^2 - \sigma^2\right)}{T_n} \sim \mathcal{N}\left(0, \Gamma(D,D) \right),
\end{equation*}

$$
\Gamma(D,D) = \mathbb{V}ar\left(D\right) = \frac{1}{T^2} \mathbb{V}ar\left(H\right) = \frac{T^2}{T^2} = 1.
$$

\Bin Hence 

$$
\frac{\sqrt{n}\left(S_n^2 - \sigma^2\right)}{T_n} \sim \mathcal{N}\left(0, 1\right).
$$

\Bin We can also write $T_n^2$ as  

\begin{equation*}
T_n^2 = S_n^2 \left(\frac{\mu_{4,n}}{ S_n^2} -  S_n^2 \right) =  S_n^2 \left(K_n(X) -  S_n^2 \right).
\end{equation*} 

\Bin Therefore 

\begin{equation*}
\frac{\sqrt{n}\left(S_n^2 - \sigma^2\right)}{S_n \left(K_n(X) -  S_n^2 \right)} \sim \mathcal{N}\left(0, 1\right).
\end{equation*}

\subsection{The Two Independent Samples Problem}

\Bin Let $X_1$, ..., $X_{n_1}$ and $Y_1$, ..., $Y_{n_2}$ be independent random variables, where $X_1$, ..., $X_{n_1}$ are iid with the functional empirical process (fep) $\mathbb{G}_{n_1}$  and $Y_1$, ..., $Y_{n_2}$ are iid with the functional empirical process (fep) $\mathbb{G}_{n_2}$. The functional empirical process $\mathbb{G}_{n_1}$ and $\mathbb{G}_{n_2}$ are assumed independent. \\

\Ni We have the notation: 

$$
\forall  \qquad i, \qquad 1\leq i\leq n_1,  \qquad \mathbb{E}X_i= \mu_x ;  \qquad  \mathbb{V}arX_i = \sigma_1^2 , \qquad m_{k,x} = \mathbb{E}X^k,  \qquad  \mu_{k,x} = \mathbb{E}\left(X- \mu_x \right)^k ,  \qquad k \geq 1,
$$

$$
\forall  \quad j, \quad 1\leq j \leq n_2,  \qquad \mathbb{E}Y_j = \mu_y ;  \qquad  \mathbb{V}arY_j = \sigma_2^2 , \qquad m_{k,y} = \mathbb{E}Y^k, \qquad \mu_{k,y} = \mathbb{E}\left(Y - \mu_y \right)^k ,  \qquad k \geq 1.
$$

\Bin We always suppose that $n_1 \geq 2$ and $n_2 \geq 2$. The main statistics are the sample mean and the sample variance for each sample. For the sample with size $n_1$, we have 

\begin{equation}
	\overline{X}_{n_1} = \frac{1}{n_1} \sum_{i=1}^{n_1} X_i,
\end{equation}

\Bin and 

\begin{equation}
	S_{n_1}^2 = \frac{1}{n_1 - 1} \sum_{i=1}^{n_1} \left(X_i - \overline{X}_{n_1} \right)^2.
\end{equation}

\Ni For the sample with size $n_2$, we have 

\begin{equation}
	\overline{Y}_{n_2} = \frac{1}{n_2} \sum_{j=1}^{n_2} Y_j,
\end{equation}

\Bin and 

\begin{equation}
	S_{n_2}^2 = \frac{1}{n_2 - 1} \sum_{j=1}^{n_1} \left(Y_j - \overline{Y}_{n_2} \right)^2.
\end{equation}

\Bin  We have  

\begin{equation*}
\overline{X}_{n_1} = \mu_x + n_1^{-1/2}\mathbb{G}_{n_1}\left( \pi_1 \right) , \quad  \pi_1(x,y) = h_1(x),
\end{equation*}

\begin{equation*}
\overline{Y}_{n_2} = \mu_y + n_2^{-1/2}\mathbb{G}_{n_2}\left( \pi_2 \right) , \quad  \pi_2(x,y) = h_1(y),
\end{equation*}

\begin{equation*}
\overline{X^2}_{n_1} = m_{2,x} + n_1^{-1/2}\mathbb{G}_{n_1}\left( \pi_3 \right) , \quad \pi_3(x,y) = h_2(x),
\end{equation*}

\begin{equation*}
\overline{Y^2}_{n_2} = m_{2,y} + n_2^{-1/2}\mathbb{G}_{n_2}\left( \pi_4 \right) , \quad \pi_4(x,y) = h_2(y).
\end{equation*}

\Bin By the delta method, we have 

\begin{equation*}
\overline{X}_{n_1}^2 = \mu_x^2 + n_1^{-1/2}\mathbb{G}_{n_1}\left(2\mu_x \pi_1 \right) + o_{\mathbb{P}}\left( n_1^{-1/2}\right)
\end{equation*}

\begin{equation*}
\overline{Y}_{n_2}^2 = \mu_y^2 + n_2^{-1/2}\mathbb{G}_{n_2}\left(2\mu_y \pi_2 \right) + o_{\mathbb{P}}\left( n_2^{-1/2}\right),
\end{equation*}

\begin{equation}
	S_{n_1}^2 = \sigma_x^2 + n_1^{-1/2} \mathbb{G}_{n_1}(H_1) + o_{\mathbb{P}}\left( n_1^{-1/2}\right),
\end{equation}

\Bin where $H_1= \pi_3 - 2 \mu_x \pi_1 = h_2 - 2\mu_x h_1$

\begin{equation}
	S_{n_2}^2 = \sigma_y^2 + n_2^{-1/2} \mathbb{G}_{n_2}(H_2) + o_{\mathbb{P}}\left( n_2^{-1/2}\right),
\end{equation}

\Bin where $H_2= \pi_4 - 2 \mu_y \pi_2 = h_2 - 2\mu_y h_1$. Now we will study the law of $\left(\frac{S_{n_1}^2}{S_{n_2}^2}\right)$. \\ 

\Bin Let $n = n_1 \wedge n_2 $.\\

\Ni We easily check $O_{\mathbb{P}}(n_j^{-1/2})=O_{\mathbb{P}}(n^{-1/2}), j =1, 2$. We also assume that $n_1 \longrightarrow + \infty$ ; $n_2 \longrightarrow + \infty$ and $(n_1)$ and $(n_2)$ are balanced between them \\

\Ni (H) $\varlimsup(\frac{n_1}{n_2}) < + \infty$ and $\varlimsup(\frac{n_2}{n_1}) < + \infty.$

\Bin We have 

\begin{equation*}
\frac{S_{n_1}^2}{S_{n_2}^2} = \frac{\sigma_1^2 + n_1^{-1/2}\mathbb{G}_{n_1}\left(H_1\right)+o_{\mathbb{P}}\left( n_1^{-1/2}\right)}{\sigma_2^2 + n_2^{-1/2}\mathbb{G}_{n_2}\left(H_2\right)+o_{\mathbb{P}}\left( n_2^{-1/2}\right)}.
\end{equation*}

\Bin By the delta method, we have 

\begin{equation}
	\left(\frac{S_{n_1}^2}{S_{n_2}^2} - \frac{\sigma_1^2}{\sigma_2^2}  \right)=  n_1^{-1/2}\mathbb{G}_{n_1}\left(\frac{1}{\sigma_2^2}H_1\right) - n_2^{-1/2}\mathbb{G}_{n_2}\left(\frac{\sigma_1^2}{\sigma_2^4}H_2\right) + o_{\mathbb{P}}\left( n^{-1/2}\right).
\end{equation}

\Bin Let 

$$
N_{1,n} = \mathbb{G}_{n_1}\left(\frac{1}{\sigma_2^2}H_1\right) = \mathcal{N}_1\left(0,T_1^2 \right) + o_{\mathbb{P}}\left(1\right),
$$

$$
N_{2,n} = \mathbb{G}_{n_2}\left(\frac{2\sigma_1^2}{\sigma_2^2}H_2\right) = - \mathcal{N}_2\left(0,T_2^2 \right) + o_{\mathbb{P}}\left(1\right),
$$

\Bin $\mathcal{N}_1$ and $\mathcal{N}_2$ are Gaussian and independent. 

$$
T_1^2 = \mathbb{V}ar\left(\frac{1}{\sigma_2^2}H_1\right) = \frac{\mu_{4,n_1} - \sigma_1^4}{\sigma_2^4},
$$

$$
T_2^2 = \mathbb{V}ar\left(\frac{\sigma_1^2}{\sigma_2^2}H_1\right) = \frac{\sigma_1^4\left(\mu_{4,n_2} - \sigma_2^4\right)}{\sigma_2^8},
$$

\Bin Let take $a_n = \sqrt{\frac{n_1n_2}{n_1T_2^2 + n_2T_1^2}}.$

\begin{theorem}  \label{gmTIS} We have

\Ni (A) For all $n_1 , n_2 \geq 2$

\begin{equation*}
	a_n\left(\frac{S_{n_1}^2}{S_{n_2}^2} - \frac{\sigma_1^2}{\sigma_2^2}\right) \sim \mathcal{N}\left(0,1\right).
\end{equation*}

\Bin (B)	For all $n_1 \geq 2$ and For all $n_2 \geq 2$ 
	
\begin{equation*}
	\sqrt{\frac{n_1n_2}{n_2T_{n_1}^2+n_1T_{n_2}^2}}\left(\frac{S_{n_1}^2}{S_{n_2}^2} - \frac{\sigma_1^2}{\sigma_2^2}\right) \sim \mathcal{N}\left(0,1\right).
\end{equation*}
\end{theorem}

\Bin\textbf{Proof of Part (A) of Theorem \ref{gmTIS}}. We recall that 

$$
\left(\frac{S_{n_1}^2}{S_{n_2}^2} - \frac{\sigma_1^2}{\sigma_2^2}\right) = n_1^{-1/2}\mathcal{N}_1\left(0,T_1^2\right) + n_1^{-1/2} o_{\mathbb{P}}(1) + n_2^{-1/2}\mathcal{N}_2\left(0,T_2^2\right) + n_2^{-1/2} o_{\mathbb{P}}(1) + o_{\mathbb{P}}\left(n^{-1/2}\right).
$$

\Bin Thus we have

\begin{align*}
a_n\left(\frac{S_{n_1}^2}{S_{n_2}^2} - \frac{\sigma_1^2}{\sigma_2^2}\right) &= \frac{a_n}{\sqrt{n_1}}\mathcal{N}_1\left(0,T_1^2\right) + \frac{a_n}{\sqrt{n_1}} o_{\mathbb{P}}(1) + \frac{a_n}{\sqrt{n_2}}\mathcal{N}_2\left(0,T_2^2\right) + \frac{a_n}{\sqrt{n_2}} o_{\mathbb{P}}(1) + a_n o_{\mathbb{P}}\left(n^{-1/2}\right) \\
&= \left(\frac{a_n}{\sqrt{n_1}}\mathcal{N}_1\left(0,T_1^2\right) + \frac{a_n}{\sqrt{n_2}}\mathcal{N}_2\left(0,T_2^2\right) \right)\\
&+ \left(\frac{a_n}{\sqrt{n_1}} o_{\mathbb{P}}(1)+\frac{a_n}{\sqrt{n_2}} o_{\mathbb{P}}(1) + a_n o_{\mathbb{P}}\left(n^{-1/2}\right) \right) \\
& = \mathcal{N}_n +  R_n,
\end{align*}

\Bin where 

$$
\mathcal{N}_n = \frac{a_n}{\sqrt{n_1}}\mathcal{N}_1\left(0,T_1^2\right) + \frac{a_n}{\sqrt{n_2}}\mathcal{N}_2\left(0,T_2^2\right),
$$

$$
R_n = \frac{a_n}{\sqrt{n_1}} o_{\mathbb{P}}(1)+\frac{a_n}{\sqrt{n_2}} o_{\mathbb{P}}(1) + a_n o_{\mathbb{P}}\left(n^{-1/2}\right) = o_{\mathbb{P}}(1).
$$

\Bin But $\mathcal{N}_n \sim \mathcal{N}\left(0, A^2\right)$ with 

$$
A^2 = \frac{a_n^2}{n_1}T_1^2 + \frac{a_n^2}{n_2}T_2^2 = \frac{n_2T_1^2}{n_1T_2^2 + n_2 T_1^2} + \frac{n_1T_2^2}{n_1T_2^2 + n_2 T_1^2} = 1.
$$

\Bin Thus

$$
a_n\left(\frac{S_{n_1}^2}{S_{n_2}^2} - \frac{\sigma_1^2}{\sigma_2^2}\right) = \mathcal{N}_n + o_{\mathbb{P}}(1) \sim \mathcal{N}\left(0, A^2 \right).
$$

\Bin Therefore 

$$
a_n\left(\frac{S_{n_1}^2}{S_{n_2}^2} - \frac{\sigma_1^2}{\sigma_2^2}\right) \sim \mathcal{N}\left(0, 1 \right)
$$

\Bin Let 

$$
	T_{n_1}^2 = \frac{\mu_{4,n_1}-S_{n_1}^4}{S_{n_2}^4}; \qquad T_{n_2}^2 = \frac{S_{n_1}^4\left( \mu_{4,n_2}-S_{n_2}^4 \right)}{S_{n_2}^8}.
$$

\Bin\textbf{Proof of Part (B) of Theorem \ref{gmTIS}}. We have 

\begin{equation*}
	T_{1,n}^2 = T_1^2 + n_1^{-1/2}\mathbb{G}_{n_1}\left(D_{11} \right) + n_2^{-1/2}\mathbb{G}_{n_2}\left(D_{12} \right) + o_{\mathbb{P}}\left( n^{-1/2}\right),
\end{equation*}

\begin{equation*}
	T_{2,n}^2 = T_2^2 + n_1^{-1/2}\mathbb{G}_{n_1}\left(D_{21} \right) + n_2^{-1/2}\mathbb{G}_{n_2}\left(D_{22} \right) + o_{\mathbb{P}}\left( n^{-1/2}\right).
\end{equation*}

\Bin We do not need to know explicit forms of the $D_{ij}$. Thus we have 

$$
	\frac{T_{1,n}^2}{n_1} = \frac{ T_1^2}{n_1} + O_{\mathbb{P}} \left( n^{-3/2}\right) =  \frac{T_1^2}{n_1}\left(1+O_{\mathbb{P}}\left(n^{-1/2}\right)\right),
$$

$$
	\frac{T_{2,n}^2}{n_2} = \frac{ T_2^2}{n_2} + O_{\mathbb{P}} \left( n^{-3/2}\right) =  \frac{T_2^2}{n_2}\left(1+O_{\mathbb{P}} \left( n^{-1/2}\right)\right).
$$

\Bin Let 
$$
	\hat{a}_n= \left(\frac{1}{\frac{T_1^2}{n_1}\left(1 + O_{\mathbb{P}}\left(n^{-1/2}\right)\right) + \frac{T_2^2}{n_2}\left(1+O_{\mathbb{P}} \left( n^{-1/2}\right)\right) }\right)^{1/2}.
$$

\Bin We recall that $a_n =\left(\frac{T_1^2}{n_1} + \frac{T_2^2}{n_2}\right)^{-1/2} $ \\

\begin{equation*}
	\hat{a}_n - a_n = \left(\hat{b}_n\right)^{-1/2} - \left(b_n\right)^{-1/2} = \left(\hat{b}_n - b_n \right) \left(\overline{b}_n \right)^{-3/2},
\end{equation*}

\Bin where $\overline{b}_n \in \left[\hat{b}_n \land b_n , \hat{b}_n \lor b_n \right]$ \\

\Bin Let $ b_n = \frac{T_1^2}{n_1} + \frac{T_2^2}{n_2}  \ \ \  and  \ \ \  \hat{b}_n = \frac{T_1^2}{n_1}\left(1+O_{\mathbb{P}}\left(n^{-1/2}\right)\right)+ \frac{T_2^2}{n_2}\left(1+O_{\mathbb{P}} \left( n^{-1/2}\right)\right).$

\Bin We have 

$$
\hat{b}_n - b_n = O_{\mathbb{P}}\left( n^{-1/2}\right),
$$

$$
\frac{b_n}{\hat{b}_n} = \frac{\frac{T_1^2}{n_1} + \frac{T_2^2}{n_2} }{\frac{T_1^2}{n_1}\left(1+O_{\mathbb{P}}\left( 1 \right)\right)+ \frac{T_2^2}{n_2}\left(1+O_{\mathbb{P}} \left(1\right)\right)}.
$$

\Bin By multiplying the numerator and denominator by $n_1 n_2$, we have 

$$
\frac{b_n}{\hat{b}_n} = \frac{n_2T_1^2 + n_1T_2^2}{n_2T_1^2\left(1+O_{\mathbb{P}}\left( 1 \right)\right)+ n_1T_2^2\left(1+O_{\mathbb{P}} \left(1\right)\right)},
$$

$$
\frac{b_n}{\hat{b}_n} = \frac{T_1^2 + \left(\frac{n_1}{n_2} \right) T_2^2}{T_1^2 \left( 1 + O_{\mathbb{P}}\left( 1 \right)\right)+ \left(\frac{n_1}{n_2} \right) T_2^2} = O_{\mathbb{P}}(1).
$$

\Bin That means $b_n \asymp \hat{b}_n$. Hence 

$$
\hat{a}_n = \left(\frac{1}{\frac{T_1^2}{n_1}\left(1+O_{\mathbb{P}}\left(n^{-1/2}\right)\right)+ \frac{T_2^2}{n_2}\left(1+O_{\mathbb{P}} \left( n^{-1/2}\right)\right) }\right)^{1/2} = \left(\frac{n_1n_2}{\frac{n_2}{n_1}T_1^2 + \frac{n_1}{n_2}T_2^2 + O_{\mathbb{P}} \left( n^{-1/2}\right)}\right)^{1/2}.
$$

\Bin Since $\frac{n_1}{n_2} \asymp 1$, so we have 

$$
\hat{a}_n =  \left(\frac{n_1n_2}{\frac{n_2}{n_1}T_1^2 + \frac{n_1}{n_2}T_2^2}\right)^{1/2} +  O_{\mathbb{P}} \left( n^{-1/2}\right),
$$

\begin{align*}
	\hat{a}_n\left(\frac{S_{n_1}^2}{S_{n_2}^2} - \frac{\sigma_1^2}{\sigma_2^2}\right) &=\sqrt{\frac{n_1n_2}{\frac{n_2}{n_1}T_1^2 + \frac{n_1}{n_2}T_2^2}} \mathbb{G}_{n_1}\left( \frac{1}{\sigma_2^2}H_1 \right) + \sqrt{\frac{n_1n_2}{\frac{n_2}{n_1}T_1^2 + \frac{n_1}{n_2}T_2^2}} \mathbb{G}_{n_2}\left( \frac{2 \sigma_1^2}{\sigma_2^2}H_2 \right) + O_{\mathbb{P}}\left(n^{-1/2} \right) \\
	&= \mathcal{N}_n(1) + \mathcal{N}_n(2)
\end{align*}

\Bin \textit{i.e.}, due to \textit{(H)}, we get

$$
\hat{a}_n  = a_n + O_{\mathbb{P}}\left( n^{-1/2} \right) = a_n + o_{\mathbb{P}}\left( 1 \right),
$$

\Bin and can conclude that 

\begin{align*}
	\hat{a}_n\left(\frac{S_{n_1}^2}{S_{n_2}^2} - \frac{\sigma_1^2}{\sigma_2^2}\right) &= a_n\left(\frac{S_{n_1}^2}{S_{n_2}^2} - \frac{\sigma_1^2}{\sigma_2^2}\right) + o_{\mathbb{P}}(1)\left(\frac{S_{n_1}^2}{S_{n_2}^2} - \frac{\sigma_1^2}{\sigma_2^2}\right)\\
	&=a_n\left(\frac{S_{n_1}^2}{S_{n_2}^2} - \frac{\sigma_1^2}{\sigma_2^2}\right) + o_{\mathbb{P}}(1).
\end{align*}

\Bin Since $a_n\left(\frac{S_{n_1}^2}{S_{n_2}^2} - \frac{\sigma_1^2}{\sigma_2^2}\right) \sim \mathcal{N}\left(0,1 \right)$, so we can conclude that 

$$
\hat{a}_n\left(\frac{S_{n_1}^2}{S_{n_2}^2} - \frac{\sigma_1^2}{\sigma_2^2}\right) \sim \mathcal{N}\left(0,1 \right).
$$

\Bin\textbf{Estimation of $\Delta m$} \\

\Bin We have 

\begin{align*}
\overline{X}_{n_1} - \overline{Y}_{n_2} - \Delta m &= n_1^{-1/2}\mathbb{G}_{n_1} \left( h_1\right) - n_2^{-1/2}\mathbb{G}_{n_2} \left( h_1\right) + o_{\mathbb{P}}\left(n^{-1/2} \right)\\
&= n_1^{-1/2}\left[ \mathcal{N}_1\left(0,\sigma_1^2 \right) + o_{\mathbb{P}}\left(1 \right)\right] + n_2^{-1/2}\left[ \mathcal{N}_2\left(0,\sigma_2^2 \right) + o_{\mathbb{P}}\left(1 \right)\right] + o_{\mathbb{P}}\left(n^{-1/2} \right) \\
&=\mathcal{N}_n + o_{\mathbb{P}}\left(n_1^{-1/2} \right) + o_{\mathbb{P}}\left(n_2^{-1/2} \right) + o_{\mathbb{P}}\left(n^{-1/2} \right)\\
&=\mathcal{N}_n + o_{\mathbb{P}}\left(n^{-1/2} \right),
\end{align*}

\Bin where 

$$
\mathcal{N}_n = n_1^{-1/2}\left[ \mathcal{N}_1\left(0,\sigma_1^2 \right)\right] + n_2^{-1/2}\left[ \mathcal{N}_2\left(0,\sigma_2^2 \right) \right] \sim \mathcal{N}\left(0,\frac{\sigma_1^2}{n_1} + \frac{\sigma_2^2}{n_2} \right),
$$

\Bin $\mathcal{N}_1$ independent with $\mathcal{N}_2$. For

$$
\tau_1^2 = \frac{\sigma_1^2}{n_1} + \frac{\sigma_2^2}{n_2} = \frac{n_2\sigma_1^2+n_1\sigma_2^2}{n_1n_2}.
$$

\Bin Then 

$$
\sqrt{\frac{n_1n_2}{n_2\sigma_1^2+n_1\sigma_2^2}}\left( \overline{X}_{n_1} - \overline{Y}_{n_2} - \Delta m\right) \sim \mathcal{N}\left(0,1 \right).
$$

\Bin Let us replace $b_n=\sqrt{\frac{n_1n_2}{n_2\sigma_1^2+n_1\sigma_2^2}}$ by $\hat{b}_n = \sqrt{\frac{n_1n_2}{n_2S_{n_1}^2+n_1S_{n_2}^2}}.$\\

\Bin Now 

$$
\hat{b}_n =\sqrt{\frac{n_1n_2}{n_2\left(\sigma_1^2 + o_{\mathbb{P}}(1) \right)+n_1\left(\sigma_2^2 + o_{\mathbb{P}}(1) \right)}}.
$$

\Bin We can easily prove, as in the letter proof, that 

$$
\hat{b}_n  = b_n + O_{\mathbb{P}}\left(n^{-1/2} \right).
$$

\Bin Hence 

$$
\hat{b}_n \left(\overline{X}_{n_1} - \overline{Y}_{n_2} - \Delta m \right) = b_n \left( n_1^{-1/2}\mathbb{G}_{n_1}\left( h_1\right)- n_2^{-1/2}\mathbb{G}_{n_2}\left( h_2\right)\right) + o_{\mathbb{P}}(1) \sim \mathcal{N}\left(0,1\right).
$$

\Bin Therefore 

$$
\sqrt{\frac{n_1n_2}{n_2S_{n_1}^2+n_1S_{n_2}^2}} \left( \overline{X}_{n_1} - \overline{Y}_{n_2} - \Delta m\right) \sim \mathcal{N}\left(0,1 \right).
$$

\section{Implementations} \label{sec_04}

\Ni Of course the Gaussian method, as an exact one, is preferable if we have the right type of data. Otherwise, we use the \textit{fep} theory as we did above. A striking fact is that we do not need to test the equality of variances in the two independent samples and by this we avoid using the \textit{Satterthwaite}.\\

\Ni From results in both situations, we draw confidence interval for estimating the means, the differences of means, the variances and the variances ratios. The results are summarized in tables below. Here, we harmonize the notation regarding the notation in the two independent samples problem. In the Gaussian case, we denoted the sample means and variances by $\{\overline{X}_n,  \ S_n^2\}$ and $\{\overline{Y}_m,  \ S_m^2\}$ with sizes $n$ and $m$ while in the general case, we used $\{\overline{X}_{n_1},  \ S_{n_1}^2\}$ and $\{\overline{Y}_{n_2},  \ S_{n_2}^2\}$ with sizes $n_1$ and $n_2$. In the sequel, the notation in the general case will prevail and in general $n_1$ and $n_2$ are used only in the two sample problem.\\

\Ni Let us recall some important definitions.\\

\begin{tabular}{lll}
\multicolumn{3}{c}{For some sample of size $n$}\\
\hline \hline
\Ni \\
$\overline{X}_n= \frac{1}{n}\sum_{j=1}^n X_j$& &$S_n^2= \frac{1}{n-1}\sum_{j=1}^n (X_j-\overline{X}_n)^2$\\
\Ni\\

$\mu_{4,n}= \frac{1}{n}\sum_{j=1}^n (X_j-\overline{X}_n)^4$ && $T_n^2=\mu_{4,n}-S_n^4$\\
& & \\

\multicolumn{3}{c}{For tow samples of size $n_1$ and $n_2$}\\
\hline \hline
\Ni\\
$T_{n_1}^2=\mu_{4,n_1}-S_{n_1}^4$&& $T_{n_2}^2=\mu_{4,n_2}-S_{n_2}^4$\\
\Ni\\
$\hat a_{n_1,n_2}^2=\frac{n_1n_2}{n_1 T_{n_2}^2+ n_2 T_{n_1}^2}$&& $\hat b_{n_1,n_2}^2=\frac{n_1n_2}{n_1 S_{n_2}^2+ n_2 S_{n_1}^2}$\\
\Ni \\

\multicolumn{3}{c}{$f=\frac{\left( \frac{S_{n}^{2}}{n}+\frac{S_{m}^{2}}{m}\right) ^{2}}{\frac{1}{n-1}\left( \frac{S_{n}^{2}}{n}\right) ^{2}+\frac{1}{m-1}\left( \frac{S_{m}^{2}}{m}\right) ^{2}}$}\\
\hline \hline
\end{tabular}

\Bin As well, we have the following quantiles: \\

\begin{tabular}{ll}
$Z_\alpha$ & quantile of the standard normal law\\
$t_\alpha(n)$ & quantile of the Student law with $n$ as $d.f$\\
$\chi_\alpha(n)$ & quantile of the Chi-square law with $n$ as $d.f$\\
$f_\alpha(n_1,n_2)$ & quantile of the Fisher law with $n_1$ and $n_2$ as $d.f$\\
\end{tabular}

\newpage
\Bin Here are the related formulas for the confidence intervals.\\

\subsection{One sample problem} \label{sec_04_01}

\begin{tabular}{ccc}
\hline \hline
Confidence interval of $m$& Inferior bound & Superior bound\\
Gaussian Data & $\overline{X}_n-\frac{S_n \ t_{1-\alpha/2}(n-1)}{\sqrt{n}}$& $\overline{X}_n+\frac{S_n \ t_{1-\alpha/2}(n-1)}{\sqrt{n}}$\\
\Ni \\

General Case & $\overline{X}_n-\frac{S_n \ z_{1-\alpha/2}}{\sqrt{n}}$& $\overline{X}_n+\frac{S_n \ z_{1-\alpha/2}}{\sqrt{n}}$\\
\hline \hline
Confidence interval of $\sigma^2$& Inferior bound & Superior bound\\
Gaussian Data & $\frac{(n-1)S_n^2}{\chi_{1-\alpha/2}^2(n-1)}$& $\frac{(n-1)S_n^2}{\chi_{\alpha/2}^2(n-1)}$\\
\Ni \\
General Case & $S_n^2 - \frac{T_n \ z_{1-\alpha/2}}{\sqrt{n}}$& $S_n^2+\frac{T_n  \ z_{1-\alpha/2}}{\sqrt{n}}$\\
\hline \hline
\end{tabular}


\Bin Here are the related formulas for the confidence intervals.\\

\subsection{Two independent sample problem} \label{sec_04_02}

\begin{tabular}{ccc}
\hline \hline
Confidence interval of $(\sigma_1/\sigma_2)^2$& Inferior bound & Superior bound\\
Gaussian Data & $\frac{S_{n_1}^2}{S_{n_2}^2 f_{1-\alpha/2}(n_1-1,n_2-1)}$ & $\frac{S_{n_1}^2}{ S_{n_2}^2 f_{\alpha/2}(n_1-1,n_2-1)}$ \\
\Ni \\
General Case & $\frac{S_{n_1}^2}{S_{n_2}^2}- z_{1-\alpha/2}/\hat a_n$& $\frac{S_{n_1}^2}{S_{n_2}^2}+ z_{1-\alpha/2}/\hat a_n$\\
\hline \hline
Confidence interval of $\Delta m$& Inferior bound & Superior bound\\
Gaussian Data & & \\
(equal variances)&$\Delta_{n_1,n_2}(X,Y)- V_{n_1,n_2}$&$\Delta_{n_1,n_2}(X,Y)+ V_{n_1,n_2}$\\
\Ni \\
  $\Delta_{n_1,n_2}(X,Y)=\overline{X}_n-\overline{Y}_n$   & \multicolumn{2}{l}{$V_{n_1,n_2}=S \ t_{1-\alpha/2}(n+m-2) \sqrt{\frac{1}{n_1}+\frac{1}{n_2}}$}\\
  \Ni \\
unequal of variances& $\Delta_{n_1,n_2}(X,Y)- W_{n_1,n_2}$&$\Delta_{n_1,n_2}(X,Y)+ W_{n_1,n_2}$\\
  & \multicolumn{2}{l}{$W_{n_1,n_2}=S \ t_{1-\alpha/2}(f) \sqrt{\frac{S_{n_1}^2}{n_1}+\frac{S_{n_2}^2}{n_2}}$}\\
 \Ni \\
General Case &$\Delta_{n_1,n_2}(X,Y)- P_{n_1,n_2}$ & $\Delta_{n_1,n_2}(X,Y)+ P_{n_1,n_2}$\\
& \multicolumn{2}{l}{ $ P_{n_1,n_2}=z_{1-\alpha/2}/\hat b_n $}\\
\hline \hline
\end{tabular}

\Ni \\

\subsection{R packages} 

\Ni \textbf{(a) Testing the exact normality}. we use the Jarque-Berra test, based on the Jarque-Berra Statistic (JBS) on data $X_j$, $1\leq j \leq n$, $n\geq 1$,

$$
J_n= n \biggr(\frac{(a_n-3)^2}{24}+\frac{b_n^2}{6}\biggr) \approx \chi_2^2,
$$

\Bin where $a_n$ are $b_n$ the empirical kurtosis and the skewness define by

$$
\overline{X}_n=\frac{1}{n}\sum_{j=1}^n X_j, \quad
a_n=\frac{\frac{1}{n}\sum_{j=1}^n (X_j-\overline{X}_n)^4}{(\frac{1}{n}\sum_{j=1}^n (X_j-\overline{X}_n)^2)^2}  \quad \text{and} \quad b_n=\frac{\frac{1}{n}\sum_{j=1}^n (X_j-\overline{X}_n)^3}{(\frac{1}{n}\sum_{j=1}^n (X_j-\overline{X}_n)^2)^{3/2}}.
$$

\Bin So, the normality accepted if the p-value $p=\mathbb{P}(\chi_2^2>J_n)$ is equal to or exceeds the level of $5\%$. Our simple R function for computing $p$ is given by the function $imhJarqueBerra(X,n)$ using the data $X$ and its size $n$ (see page \pageref{NTJB}).\\

\Bin \textbf{(b) Tests of one sample}.\\

\Ni Our function \textit{imhSimpleTestZ,nn,alpha1,alpha2,dg)} (see \pageref{OSP}) provides all results in the table in Subsection \ref{sec_04_01}, page \pageref{sec_04_01}  above.\\

\Bin \textbf{(b) Two samples Problem}.\\

\Ni Our function \textit{imhTwoSamplesTest(Z1,Z2,n1,n2,alpha1,alpha2,dg)} (see \pageref{OSP}) provides all results in 
in the tablein Subsection \ref{sec_04_02}, page \pageref{sec_04_02}  above

\Ni these function are explained as well as their application in the above mentioned pages.

\section{Data-driven applications and simulation} \label{sec_05}

\Ni We are going to apply our both methods and the Gaussian method to a set of real-life data or to simulated data. Studies are called \textit{scenarios}. Each scenario commented and general comments and conclusions are proposed at the end.

\subsection{Simulated Gaussian data of small size (n=9)}\label{sec_05_01}

\Ni Let us begin by comparing the two methods on Gaussian data with small sizes around $n=9$. We generate $X=rnorm(n,3,2)$ with $m=3$ and 
$\sigma^2=4$. We apply our R function \textit{imhSimpleTest(X,n,$\alpha1$,$\alpha2$, dg)}
[with $\alpha1=0.05$, $\alpha2=0.1$, $dg=2$] to get the following results:\\

\Bin

\begin{tabular}{ccc}
\hline \hline
Confidence interval of $m$& Inferior bound & Superior bound\\
Gaussian Data & $1.9$& $4.3$\\
General case & $2.1$& 4.1$$\\
\hline \hline
Confidence interval of $\sigma^2$& Inferior bound & Superior bound\\
Gaussian Data & $1.3$& $7.3$\\
General case & $0.54$& $4.5$\\
\hline \hline
\end{tabular}

\Bin After a significant number replications, we get similar outputs with the following regards:\\

\Ni (1) Generally the mean estimation is more precise for the General case.\\
\Ni (2) Generally the variance  estimation is more precise for the General case,. However the inferior can be negative in the general case and we get an confidence interval of the form $[0, \ c]$ with $c>0$ always less than it counterpart for the Gaussian Case.\\

\Ni \textbf{Partial conclusion [1]}. Even with weak sizes as for $n=9$, the general case beats the Gaussian case for the mean estimation. As to the variance estimation, the general has the drawback to possibly present a negative inferior bound.

\subsection{Data-driven Gaussian Data for small size (n=9)} \label{sec_05_02}

\Ni We may remake the same application to the data-driven application with the \textit{Sheffé data}.  Let us consider $A1$ as the data in Batch 1 for machine A (first in the the data) and $B1$ as the data in Batch 1 for machine B (fourth line in the data). Comparing the means $A1$ and (B1) is performed by studying the the variable $AB=A1-B1$. The Jarque-Bera test for $AB$ gives: $JBS=0.13$ and $p-value: 93.647\%$.  The normality of the data is hence assumed. The results of the one sample study are summarized in the table below. Here the confidence intervals for the means use a $5\%$-level of significance As for the variance a $10\%$-level of significance is used in the Gaussian case and a $5\%$-level of significance is used for the general case. We apply our R function \textit{imhSimpleTest(AB,nn,$\alpha1$,$\alpha2$, dg)}
[with $nn=9$, $\alpha1=0.05$, $\alpha2=0.1$, $dg=2$]. Here is an output for one replication.\\

\Bin

\begin{tabular}{ccc}
\hline \hline
Confidence interval of $m$& Inferior bound & Superior bound\\
Gaussian Data & $0.11$& $0.47$\\
General case & $-1.9$& $2$\\
\hline \hline
Confidence interval of $\sigma^2$& Inferior bound & Superior bound\\
Gaussian Data & $0.028$& $0.16$\\
General case & $-1.9$& $2$\\
\hline \hline
\end{tabular}
 
\Bin \textbf{In this present case}, the better performance of Machine $A1$ over Machine $B1$ is validated by the marge $\Delta m\geq 0.11$ trough the Gaussian exact distribution method while the general method fails to do it. We already recommended the application of the method from $n\approx 15$.\\

\Bin \textbf{Partial conclusion [2]}. Here we are in the wrong $5\%$. The results for the general case present not good inferior bounds in the mean and the variance estimation. We get the following recommendation: (R1) For sizes equal to less $9$, we recommend to be careful for the application of the general case because not efficient inferior bounds.\\

\subsection{What size make the application of General method reasonable?} \label{sec_05_03}

\Ni Form several experiments, we recommend the application of the general case for sizes from $n=15$. Here are outputs for $X=rnorm(n,3,2)$ with $n=29$, $m=3$ and $\sigma^2=4$, \textit{imhSimpleTest(X,n,$\alpha1$,$\alpha2$, dg)} [with $\alpha1=0.05$, $\alpha2=0.1$, $dg=2$] to get the following results:\\

\Bin

\begin{tabular}{ccc}
\hline \hline
Confidence interval of $m$& Inferior bound & Superior bound\\
Gaussian Data & $2.2$& $4$\\
General case & $2.3$& $3.9$\\
\hline \hline
Confidence interval of $\sigma^2$& Inferior bound & Superior bound\\
Gaussian Data & $2.2$& $6.5$\\
General case & $2$& $4.9$\\
\hline \hline
\end{tabular}

\Bin \textbf{Partial conclusion [3]}. All results for the General case are . Still, the inferior bound for the variance estimation, evecn it is correct, is a little underestimated compared to the Gaussian case.\\

\subsection{Application to Senegalese data} \label{sec_05_04}

\Ni We apply the methods to randomly picked data of sizs $n=50$ from \textit{dakar1}, \textit{dakar2}, \textit{diour1}, \textit{diour2}. The Jarque-Berra test gives almost null p-values $(<10^{-9})$ for all of them. SInce the sizes are big enough (nn=50), the \textit{QQ-plots} are efficient to check the nor normality. Ces \textit{QQ-plots} can be found at page \pageref{qqplots_esam_12}.\\

\Ni We have the following outputs for \textit{dakar1} then for \textit{dakar2}:\\

\Ni For \textit{dakar1}:
\Bin

\begin{tabular}{ccc}
\hline \hline
Confidence interval of $m$& Inferior bound & Superior bound\\
Gaussian Data & $190 746$& $786 611$\\
General Case & $200 000$& $779 256$\\
\hline \hline
Confidence interval of $\sigma^2$& Inferior bound & Superior bound\\
Gaussian Data & $8.1 \times 10^{11} $& $1.6 \times 10^{12}$\\
General Case & $1.1 \times 10^{11}$& $1.1 \times 10^{12}$\\
\hline \hline
\end{tabular}

\Bin For \textit{dakar2}.\\
 
\Ni
\begin{tabular}{ccc}
\hline \hline
Confidence interval of $m$& Inferior bound & Superior bound\\
Gaussian Data & $542 666$& $990 509$\\
General Case & $548 194$& $984 981$\\
\hline \hline
Confidence interval of $\sigma^2$& Inferior bound & Superior bound\\
Gaussian Data & $4.6 \times 10^{11}$& $9 \times 10^{11}$\\
General Case & $6.2 \times 10^{11}$& $6.2 \times 10^{12}$\\
\hline \hline
\end{tabular}

\Bin \textbf{Partial conclusion [4]}. In this non-Gaussian analysis, the confidence intervals for the mean have smaller amplitudes. The results are thus more precise, making the methods we introduced reliable.\\

\subsection{The two samples problem} \label{sec_05_06}

\Ni Here we expected to see the same conclusion as in the general case. We will see how things really are after some scenarii: 
We apply our function \textit{imhTwoSamplesTest(Z1,Z2,n1,n2,$\alpha1$,$\alpha2$, dg)} [with $\alpha1=0.05$, $\alpha2=0.1$, $dg=2$ to study the estimation of $\Delta m$.\\

\Ni \textbf{(a) Comparison between performances of $A1$ and $B1$ in Scheff\'e's data}.\\

\Ni Here, we use $nn=9$, $Z1=A1$, $Z2=B1$, $dg=4$. We get\\

\begin{tabular}{ccc}
\hline \hline
Confidence interval of $(\sigma_1/\sigma_2)^2$& Inferior bound & Superior bound\\
Gaussian Data & $0.34$& $4.4$\\
General Case & $1.254$& $1.285$\\
\hline \hline
Confidence interval of $\Delta m$& Inferior bound & Superior bound\\
Gaussian Data & & \\
(equal variances)&$0.093$&$0.48$\\
unequal of variances& $0.093$&$0.48$\\
General Case &$0.11$ & $0.46$\\
\hline \hline
\end{tabular}

\Bin

\Ni \textbf{(b) Comparison between Gaussian Data for $n1$ and $n2$ small}.\\

\Ni Here, we use $nn=9$, $Z1=rnorm(nn,4, \sqrt{6})$, $Z2=rnorm(nn,1, \sqrt{2}$), $dg=4$. We get \\

\begin{tabular}{ccc}
\hline \hline
Confidence interval of $(\sigma_1/\sigma_2)^2$& Inferior bound & Superior bound\\
Gaussian Data & $0.33$& $4.32$\\
General Case & $-0.57$& $3.034$\\
\hline \hline
Confidence interval of $\Delta m$& Inferior bound & Superior bound\\
Gaussian Data & & \\
(equal variances)&$0.76$&$4.26$\\
unequal of variances& $0.76$&$4.26$\\
General Case &$0.89$ & $4.13$\\
\hline \hline
\end{tabular}

\Bin 

\Ni \textbf{(c) Comparison between Gaussian Data for $n1\geq 15$ and $n2\geq 15$ small}.\\

\Ni We only change $nn=15$ in the precedent case.

\begin{tabular}{ccc}
\hline \hline
Confidence interval of $(\sigma_1/\sigma_2)^2$& Inferior bound & Superior bound\\
Gaussian Data & $1.35$& $8.69$\\
General Case & $-0.91$& $7.82$\\
\hline \hline
Confidence interval of $\Delta m$& Inferior bound & Superior bound\\
Gaussian Data & & \\
(equal variances)&$1,63$&$4.25$\\
unequal of variances& $1.63$&$4.25$\\
General Case &$1.23$ & $4.18$\\
\hline \hline
\end{tabular}

\Bin \textbf{Partial conclusion [5]}. From scenarii (a), (b) and (c) above, we have three striking remarks about the general case, two of them being positive and the other negative. (i) The estimation of $\Delta m$ is always more precise with the general method. (ii) The estimation through the general case does not care about the equality of the variances. (iii) However, the inferior bound for the estimation of the variance ratio can take the value zero. At the same time, the superior bound is less than in the case of Gaussian data. Since, the estimation of $\Delta m$ does not depend of the variances equality, this does not affect the efficiency in the method.\\

\Ni \textbf{(d) Comparison between non-Gaussian data in Esam Senegalese data}.\\

\Ni \textbf{(d1) Evolution of area's income between two periods}.\\

\Ni We use nn=50, 
imhTwoSamplesTest(dak2E,dak1E,nn,nn,0.05,0.1,4).\\

\begin{tabular}{ccc}
\hline \hline
Confidence interval of $\Delta m$& Inferior bound & Superior bound\\
Gaussian Data & & \\
(equal variances)&$-291309$&$164352$\\
unequal of variances& $-291309$&$164352$\\
General Case &$-288496$ & $161354$\\
\hline \hline
\end{tabular}

\Bin

\Ni \textbf{(d2) Geographical discrepancy income between Dakar and Diourbel in 1996}

\begin{tabular}{ccc}
\hline \hline
Confidence interval of $\Delta m$& Inferior bound & Superior bound\\
Gaussian Data & & \\
(equal variances)&$214 102$&$587 024$\\
unequal of variances&$214 102$&$587 024$\\
General Case &$216 405$ & $584v722$\\
\hline \hline
\end{tabular}

\Bin 

\Ni \textbf{(d3) Geographical discrepancy income between in 2000}

\begin{tabular}{ccc}
\hline \hline
Confidence interval of $\Delta m$& Inferior bound & Superior bound\\
Gaussian Data & & \\
(equal variances)&$157 976$&$438 435$\\
unequal of variances&$157 976$&$438 435$\\
General case &$159 708$ & $436 705$\\
\hline \hline
\end{tabular}

\Bin \textbf{Partial conclusion [6]}. In scenarios (d1), (d2) and (d3), we applied our method to income Senegalese data that are not Gaussian.
In general, as already noticed, the confidence intervals for estimated $\Delta m$ are more precise with smaller amplitude. By using them, say that:\\

\Ni (i) the income does not change from 1996 to 2000. We see this for Dakar. We did not report the corresponding outputs for Diourbel since the conclusion is still the same.\\

\Ni (ii) For each period, the income is far higher in Dakar than in Diourbel with significant thresholds of $216 405$ in 1996
and $159 708$. The good thing is that the discrepancy diminished by $26,19\%$.

\section{Conclusion} \label{sec_06}
\Ni The new method, called the General Samples Problem, is based on asymptotic methods, specially on the striking results of the \textit{functional empirical process (fep)}. Almost for samples' sizes of $n=10$, the results are almost equivalent to those in the Gaussian case. For greater sizes, the estimation of the mean difference is very precise regardless of the equality or inequality of variances. We applied the results to Senegalese data which are not Gaussian as checked by the use of Jarque-Bera test. We recommend that method for sizes around or greater that $n=15$. However, the method requires the finiteness of the fourth order moment. We have to be cautious about this when dealing with heavy tails.\\

\newpage
\Ni \textbf{APPENDIX DATA}\\ \label{}

\Ni \textbf{Data from \cite{ScheffeAV}, page. 290}\\ \label{scheffe}

\begin{figure}[htbp]
	\centering
		\includegraphics[width=0.98\textwidth]{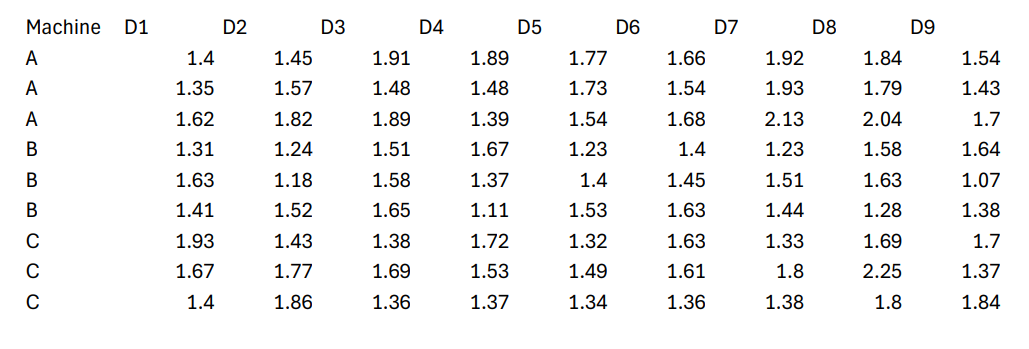}
	\label{data_scheffe}
\end{figure}

\Ni \textbf{Income Data of Dakar 1}.\label{dak1}\\

\bigskip

\begin{tabular}{c|c|c|c|c|c}
\hline \hline
212860.9 	& 267010.7 	& 258080.4 	& 578540.3  &244696.2 	& 215546.2\\
290591.1	& 304817.6  &291419.1  	&440473.7  	&138962.8  	&145774.7\\
261535.4 	&1182688.3  &231771.8  	&108884.2  	&583703.4  	&340638.7\\
143220.2 	& 361730.6  &488192.8  	&232069.2  	&599704.8  	&194193.3\\
320658.9 	& 167081.5  &236670.5 	&1027139.7  &226607.0  	&567213.9\\
714605.4  &107189.7  	&160302.3  	&363402.8  	&761425.0  	&466297.9\\
179790.5  &179206.9 	&7591873.8  &257280.9  	&247732.1  	&271506.2\\
242673.6  &495594.4  &204820.5		&  182907.2	&  314469.1 &390925.7\\
388434.4  &251026.1  &        		&         	&            &\\
\hline \hline
\end{tabular}

\newpage

\Ni \textbf{Income Data of Dakar 2}.  \label{dak2}

\bigskip

\begin{tabular}{c|c|c|c|c|c}
\hline \hline
2326347.1&202446.1&497640.6 &1505369.5&206145.4&670007.1\\
688580.6 &3800902.5&456837.8&772329.3 &1268005.4&944911.9\\
579565.7&345986.3&605220.4&275933.6&664973.9&345986.3\\
1443230.0&3111786.7&190088.7&400339.4 &1440574.8&456007.6\\
204802.6&400339.4&901053.5&184492.8&900889.4&324556.9\\
612592.7&603949.7&705670.0&255847.5&711869.7 &2385253.6\\
129207.5&354822.7&175583.9&206145.4&532207.0&224469.6\\
364105.7&609470.4&751249.5&129207.5 2&441979.0&328765.1\\
374895.0&316734.3\\
\hline \hline
\end{tabular}

\bigskip
\Ni \textbf{Income Data of Diourbel 1}. \label{dio1}\\

\bigskip

\begin{tabular}{c|c|c|c|c|c}
\hline \hline
165087.11&191031.56&123808.57&396346.90&82781.90&519556.29\\
146485.27&144961.70&122684.94 &146399.19&349398.28&153086.53\\
292284.34&160460.77&194542.64& 82656.72&214976.94&151839.94\\
160460.77&104743.29&662028.92&132762.82&91387.36&211374.42\\
135755.33&144137.22& 75249.22& 87559.66&163498.36& 89995.23\\
125835.76&300345.97&257382.04&215629.97&109558.28&214976.94\\
87289.32&233952.60&104693.64&197164.40&343203.01&385182.29\\
82781.90&658761.83&215069.65&117829.49&132192.84&95234.29\\
216132.15 &396346.90 \\
\hline \hline
\end{tabular}

\bigskip
\Ni \textbf{Income Data of Diourbel 2}. \label{dio2}\\

\bigskip
\begin{tabular}{c|c|c|c|c|c|}
\hline \hline
339293.85 &	199794.97 &	167231.82	&	317826.88 &	473868.40 	&	440738.61\\
150857.01 &	205082.61 &	480722.21 &	160956.28 &	313255.46 	&	267165.79\\
205322.67 &	268027.99 &	78512.00 	&242960.37 	&152158.57 		&182687.95\\
396531.52 &	156383.34 &	205082.61 &	393941.15 &	810344.85		&99643.59\\
107577.92 &	242895.53 &	299596.63 &	138594.52 &	205852.70		&94937.99\\
190404.38 &	317826.88 &	 97794.50 &	192905.57 &	167724.34 	&777178.29\\
190377.06 &	366065.69 &	152322.68 &	186260.20 &	474140.29 	&222936.34\\
325813.46 &	138305.38 &	514067.87	&	77420.32 	&211870.89 	 	&189376.45\\
223896.62 &345868.96 	&						&						&							&\\
\hline \hline
\end{tabular}

\newpage
\Ni \textbf{qq-plots against normal law for data from ESAM}.

\begin{figure}[ht]
	\centering
		\includegraphics[width=0.99\textwidth]{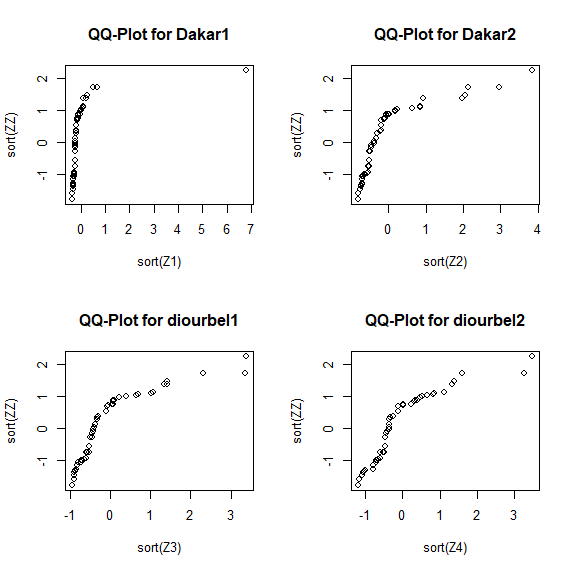}
	\caption{QQ-plots of data from ESAM against Normal law}
	\label{qqplots_esam_12}
\end{figure}

\newpage
\Ni \textbf{APPENDIX PACKAGES}\\

\Ni  \textbf{1. Normality Test of Jarque-Berra} \label{NTJB}\\

\begin{lstlisting}
# Ce package requires the data vector V of size at least one
# the size TT of the data
# Sortie: display skewness, kurtosis, jarque-Bera Statistics
# and pv-alue of the test.
# At he first running, initialize as follows
TT=15
V <- rnorm(TT,0,1)

imhJarqueBerra <- function(V, XX){
	va=sqrt(mean((V-mean(V))^2))
	va
	a0n=(mean((V-mean(V))^4)/(va^4))
	an=(a0n-3)^2
	b0n=(mean((V-mean(V))^3)/(va^3))
	bn=b0n^2
	jbt=taille*((an/24)+(bn/6))
	ret=numeric(4)
	ret[1]=paste("skewness:", b0n)
	ret[2]=paste("kurtosis:", a0n)
	ret[3]=paste("Jarque Berra Statistic:", jbt)
	ret[4]=paste("pa-value: ", format(100*(1-pchisq(jbt,2)),digits=5),"%")
	return(noquote(ret))
}
# Apply the function as : imhJarqueBerra(X,nn)
#Thar gives the p-values for data X of size nn
\end{lstlisting}

\Bin \textbf{2. A display function} \label{ADPSP}\\
 
\begin{lstlisting}
# This function display the the number dg of digits of string zz 
# with  no quotes. The limitation of the number of digits
# works only of zz is originally a number
(zz=12.2213)
imhAff2 <- function(zz,dg){
         aff=noquote(format(zz,digits=dg))
		 return(aff)
}
 #Apply the function as : imhAff2(335.98673,5)
\end{lstlisting}

\newpage
\Bin \textbf{2. One sample problem} \label{OSP}\\

\begin{lstlisting}
# This function has as arguments: a data set Z of size nn,
# alpha1: seuil of significance of estimating means or ther differences
# alpha2: seuil of significance of testing variances of their rations
# dg=2: number of digits for the display of the results
# Initialize as follows:
nn=30
m=2
s=4
Z <- rnorm(nn,m,s)
alpha1=0.05
alpha2=.1

imhSimpleTest <- function(Z,nn,alpha1,alpha2,dg){
guenne=c(1:14)
(tc=qt(1-alpha1/2,nn-1))
(uc1=qnorm(1-alpha1/2,0,1))
(uc2=qnorm(1-alpha2/2,0,1))
moy=mean(Z)
sigma=sd(Z)
amp=sigma*tc/sqrt(nn)
ampGm=sigma*uc1/sqrt(nn)
minf=moy-amp
msup=moy+amp
mginf=moy-ampGm
mgsup=moy+ampGm
mu4= mean((Z-mean(Z))^4)
(TnC=mu4-(sigma^4))
ampGsigma=(sigma^2)- sqrt(TnC)*uc1/sqrt(nn)
## Case general : alpha=0.05
sigInfG=(sigma^2)-ampGsigma
sigSupG=(sigma^2)+ampGsigma

(sigInf=(nn-1)*(sigma^2)/qchisq(1-alpha2/2,nn-1))
(sigSup=(nn-1)*(sigma^2)/qchisq(alpha2/2,nn-1))


guenne[1]="======================"
guenne[2]=paste('Estimation of the mean')
guenne[3]=paste('Gaussian case')
guenne[4]=paste(imhAff2('Mean:',dg),imhAff2(moy,dg),imhAff2('/alpha',dg),
imhAff2(alpha1,dg),imhAff2('/BiCI:',dg),imhAff2(minf,dg),
imhAff2('/BsCI:',dg),imhAff2(msup,dg))

guenne[5]="======================"
guenne[6]=paste('General Case')
guenne[7]=paste(imhAff2('Mean:',dg),imhAff2(moy,dg),
imhAff2('/alpha',dg),imhAff2(alpha1,dg),imhAff2('/BiCI:',dg),
imhAff2(mginf,dg), imhAff2('/BsCI:',dg),imhAff2(mgsup,dg))

guenne[8]="======================"
guenne[9]=paste('Estimation of the variance')
guenne[10]=paste('Gaussian case')
guenne[11]=paste(imhAff2('Variance:',dg),imhAff2(sigma^2,dg),
imhAff2('/alpha',dg), imhAff2(alpha2,dg),imhAff2('/BiCI:',dg),
imhAff2(sigInf,dg),imhAff2('/BsCI:',dg),imhAff2(sigSup,dg))

guenne[12]="======================"
guenne[13]=paste('General Case')
guenne[14]=paste(imhAff2('Variance:',dg),imhAff2(sigma^2,dg),
imhAff2('/alpha',dg),imhAff2(alpha1,dg),imhAff2('/BiCI:',dg),
imhAff2(sigInfG,dg),imhAff2('/BsCI:',dg),imhAff2(sigSupG,dg))
	
return(noquote(guenne))
}

#Finally, apply the function as:
# imhSimpleTestZ,nn,alpha1,alpha2,dg)
\end{lstlisting}

\newpage

\Bin \textbf{2. Two samples problem} \label{TSP}\\

\begin{lstlisting}
# This function has as arguments: two data sets Z1 and Z2 
#of size n1 and n2,
# alpha1: seuil of significance of estimating means 
#or ther differences
# alpha2: seuil of significance of testing variances
# of their rations
# dg=2: number of digits for the display of the results
# Initialize as follows:
#======
n1=25
n2=25
m1=2
m2=5
sigma1C=3
sigma2C=7
Z1=rnorm(n1, m1, sigma1C)
Z2=rnorm(n2, m2, sigma1C)
Dm=m1-m2

imhTwoSamplesTest <- function(Z1,Z2,n1,n2,alpha1,alpha2,dg){
guenne=c(1:16)
Dmoy=mean(Z1)-mean(Z2)
sigma1EC=var(Z1)
sigma2EC=var(Z2)
RV=(sigma1C/sigma2C)
RVE=(sigma1EC/sigma2EC)
de=((sigma1EC/n1)+(sigma2EC/n2))^2
num=((sigma1EC/n1)^2/(n1-1))+((sigma2EC/n2)^2/(n2-1))
(fs=de/num)
(tcf=qt(1-alpha1/2,fs))
(uc1=qnorm(1-alpha1/2,0,1))
mu4E1= mean((Z1-mean(Z1))^4)
mu4E2= mean((Z2-mean(Z2))^4)
T1C=mu4E1-(sigma1EC^2)
T2C=mu4E2-(sigma2EC^2)
achap=sqrt((n1*n2)/((n1*T2C)+(n2*T1C)))
bchap=sqrt(n1*n2/((n1*sigma2EC)+(n2*sigma1EC)))

#sigmaRatio
srInf=(sigma1EC/sigma2EC)/qf(1-alpha2/2,n1-1,n2-2)
srSup=(sigma1EC/sigma2EC)/qf(alpha2/2,n1-1,n2-2)
srInfG=(sigma1EC/sigma2EC) - (uc1/achap)
srSupG=(sigma1EC/sigma2EC) + (uc1/achap)
(sigma1EC/sigma2EC)
#sqrt(((T1C/n1)+(T2C/n2)))
(uc1/achap)
uc1*achap
#Global S
gsc=((n1-1)*sigma1EC+(n2-1)*sigma2EC)/(n1+n2-2)
gs=sqrt(gsc)

quotEV= sqrt((1/n1)+(1/n2))
quotUV= sqrt((sigma1EC/n1)+(sigma2EC/n2))
 
DiffAmpN= gs*qt(1-alpha1/2,n1+n2-2)*quotEV
biN=Dmoy-DiffAmpN
bsN=Dmoy+DiffAmpN

DiffAmpAp= qt(1-alpha1/2,n1+n2-2)*quotUV
biA=Dmoy-DiffAmpAp
bsA=Dmoy+DiffAmpAp

DiffAmpG= uc1/bchap
biG=Dmoy-DiffAmpG
bsG=Dmoy+DiffAmpG

guenne[1]="======================"
guenne[2]=paste('Estimation of the ratio of variance')
guenne[3]=paste('Gaussian case')
guenne[4]=paste(imhAff2('Ratio:',dg),imhAff2('/alpha',dg),
imhAff2(alpha2,dg),
imhAff2('/BiCI:',dg),imhAff2(srInf,dg),imhAff2('/BsCI:',dg),
imhAff2(srSup,dg))

guenne[5]="======================"
guenne[6]=paste('General Case')
guenne[7]=paste(imhAff2('Ratio:',dg),imhAff2('/alpha',dg),
imhAff2(alpha1,dg),imhAff2('/BiCI:',dg),imhAff2(srInfG,dg),
imhAff2('/BsCI:',dg), imhAff2(srSupG,dg))

guenne[8]="======================"
guenne[9]=paste('Estimation of the mean difference')
guenne[10]=paste('Gaussian case with equality of variances')
guenne[11]=paste(imhAff2('Diff:',dg),imhAff2(Dm,dg),
imhAff2('/alpha',dg), imhAff2(alpha1,dg),imhAff2('/BiCI:',dg),
imhAff2(biN,dg), imhAff2('/BsCI:',dg),imhAff2(bsN,dg))

guenne[12]="======================"
guenne[13]=paste('Gaussian case with unequality of variances')
guenne[14]=paste(imhAff2('Diff:',dg),imhAff2('/alpha',dg),
imhAff2(alpha1,dg), imhAff2('/BiCI:',dg),imhAff2(biA,dg),
imhAff2('/BsCI:',dg),imhAff2(bsA,dg))

guenne[15]="======================"
guenne[16]=paste('General Case')
guenne[17]=paste(imhAff2('Diff:',dg),imhAff2('/alpha',dg),
imhAff2(alpha1,dg), imhAff2('/BiCI:',dg),imhAff2(biG,dg),
imhAff2('/BsCI:',dg),imhAff2(bsG,dg))
	return(noquote(guenne))
}

# Use the function as :
# imhTwoSamplesTest(Z1,Z2,n1,n2,alpha1,alpha2,dg)
\end{lstlisting}

\end{document}